\documentclass[apj,numberedappendix]{emulateapj}
\usepackage{graphics,color}
\usepackage{amssymb,txfonts}
\usepackage{epsfig,natbib}
\newcommand{\black}{\color{black}}
\newcommand{\jt}{\black}

\DeclareMathAlphabet{\mathitbf}{OML}{cmm}{b}{it}
\DeclareMathAlphabet{\mathf}{OML}{cmm}{c}{sl}
\renewcommand{\vec}[1]{ {\mathitbf #1}}
\newcommand{\Bvec}{\vec B}
\newcommand{\Jvec}{\vec J}
\newcommand{\Jl}{\vec \Jvec_{l}}
\newcommand{\Jlabs}{\mathit J_l}

\newcommand{\Jabs}{\mathit J}
\newcommand{\Blos}{\Bvec_{\rm LOS}}
\newcommand{\Btrans}{\Bvec_t}

\newcommand{\Bobs}{\vec B_{\rm obs}}
\newcommand{\Bpot}{\vec B_{\rm pot}}
\newcommand{\Bcurr}{\vec B_{\rm curr}}
\newcommand{\Bnlff}{\vec B_{\rm nlff}}

\newcommand{\Bn}{{\mathf B_{n}}}
\newcommand{\Bnnlff}{{\mathf B_{n}^{\rm nlff}}}
\newcommand{\Bx}{{\mathf B_x}}
\newcommand{\By}{{\mathf B_y}}

\newcommand{\Babs}{\mathit B}

\newcommand{\Btabs}{\mathit B_t}
\newcommand{\Blabs}{\mathit B_l}
\newcommand{\Blabsvsmme}{\Blabs^{\rm VSM-ME}}
\newcommand{\Blabsvsmql}{\Blabs^{\rm VSM-QL}}
\newcommand{\Blabshmi}{\Blabs^{\rm HMI}}

\newcommand{\Btabsvsmme}{\Btabs^{\rm VSM-ME}}

\newcommand{\Btabshmi}{\Btabs^{\rm HMI}}
\newcommand{\cosxi}{\cos(\xi)}
\newcommand{\Em}{E_{\rm m}}
\newcommand{\Evir}{E_{\rm vir}}
\newcommand{\Enlff}{E_{\rm nlff}}
\newcommand{\Epot}{E_{\rm pot}}
\newcommand{\Evirhmi}{\Evir^{\rm HMI}}
\newcommand{\Evirvsmme}{\Evir^{\rm VSM-ME}}
\newcommand{\Evirvsmql}{\Evir^{\rm VSM-QL}}
\newcommand{\Efree}{\Delta E_{\rm m}}
\newcommand{\pmag}{p_{\rm m}}
\newcommand{\pmaghmi}{p_{\rm m}^{\rm HMI}}
\newcommand{\pmagvsmme}{p_{\rm m}^{\rm VSM-ME}}
\newcommand{\pmagvsmql}{p_{\rm m}^{\rm VSM-QL}}
\newcommand{\fsign}{\phi_{\rm m}}

\newcommand{\fu}{|\phi_{\rm m}|}

\newcommand{\fuvsmme}{|\phi^{\rm VSM-ME}_{\rm m}|}
\newcommand{\fuvsmql}{|\phi^{\rm VSM-QL}_{\rm m}|}
\newcommand{\fuhmi}{|\phi^{\rm HMI}_{\rm m}|}

\newcommand{\eforce}{\epsilon_{\rm force}}
\newcommand{\etorque}{\epsilon_{\rm torque}}
\newcommand{\Wvec}{\vec W}

\newcommand{\rvec}{\vec r}

\newcommand{\Lforce}{\mathcal L_{\rm force}}
\newcommand{\Ldiv}{\mathcal L_{\rm div}}
\newcommand{\Tj}{\theta_{\Jvec}}

\newcommand{\Lopt}{\mathcal L_{\rm opt}}
\newcommand{\wforce}{w_{\rm f}}
\newcommand{\wdiv}{w_{\rm d}}
\newcommand{\vol}{\mathcal V}
\newcommand{\surf}{\mathcal S}
\newcommand{\intv}{\int_\vol}
\newcommand{\dvol}{~{\rm d}\vol}
\newcommand{\dsurf}{~{\rm d}\surf}
\newcommand{\ints}{\int_\surf}
\newcommand{\volsub}{\vol_{\rm sub}}
\newcommand{\volel}{\partial \mathcal{V}}
\newcommand{\volels}{\volel_{\rm sub}}
\newcommand{\wtrans}{\rm w_{trans}}
\newcommand{\wlos}{\rm w_{\rm LOS}}

\newcommand{\Cpot}{\rm C_{\rm pot}}
\newcommand{\Ccurr}{\rm C_{\rm curr}}
\newcommand{\Cnlff}{\rm C_{\rm nlff}}

\newcommand{\km}{{\rm\,km}}
\newcommand{\G}{{\rm\,G}}
\newcommand{\kG}{{\rm\,kG}}
\newcommand{\Mx}{{\rm\,Mx}}
\newcommand{\J}{{\rm\,J}}
\newcommand{\A}{{\rm\,A}}

\newcommand{\mAm}{{\rm\,mAm^{-2}}}
\newcommand{\Mm}{{\rm\,Mm}}
\newcommand{\degree}{^\circ}
\newcommand{\ie}{i.\,e.}
\newcommand{\eg}{e.\,g.}
\newcommand{\etal}{et al.}

\slugcomment{The Astronomical Journal; Accepted May 11, 2012}
\shorttitle{Force-free modeling using HMI and VSM data}
\shortauthors{Thalmann \etal}

\begin{document}

\title{Nonlinear force-free field modeling of a solar active region\\
using SDO/HMI and SOLIS/VSM data}

\author{J. K. Thalmann\altaffilmark{1}, A. Pietarila\altaffilmark{2}, X. Sun\altaffilmark{3} and T. Wiegelmann\altaffilmark{1}}
\email{thalmann@mps.mpg.de}
\altaffiltext{1}{Max-Plank-Institut f\"ur Sonnensystemforschung, Max-Planck-Str. 2, 37191 Katlenburg-Lindau, Germany}
\altaffiltext{2}{National Solar Observatory, 950 N. Cherry Avenue, Tucson, AZ 85719, USA}
\altaffiltext{3}{W.\,W.\,Hansen Experimental Physics Laboratory, Stanford University, Stanford, CA 94305, USA}

\begin{abstract}
We use SDO/HMI and SOLIS/VSM photospheric magnetic field measurements to model the force-free coronal field above a solar active region, assuming magnetic forces to dominate. We take measurement uncertainties caused by, \eg, noise and the particular inversion technique into account.  After searching for the optimum modeling parameters for the particular data sets, we compare the resulting nonlinear force-free model fields. We show the degree of agreement of the coronal field reconstructions from the different data sources by comparing the relative free energy content, the vertical distribution of the magnetic pressure and the vertically integrated current density. Though the longitudinal and transverse magnetic flux measured by the VSM and HMI is clearly different, we find considerable similarities in the modeled fields. This indicates the robustness of the algorithm we use to calculate the nonlinear force-free fields against differences and deficiencies of the photospheric vector maps used as an input. We also depict how much the absolute values of  the total force-free, virial and the free magnetic energy differ and how the orientation of the longitudinal and transverse components of the HMI- and VSM-based model volumes compare to each other.
\end{abstract}


\section{Introduction}

Presently, no measurements of the magnetic field vector in the outer solar atmosphere, the corona, are routinely performed. Nevertheless, its magnetic structuring is of great interest and an active field of research. To overcome the unavailability of coronal magnetic field measurements, photospheric magnetic field vector data can be used to reconstruct the coronal field. Photospheric vector data are obtained 
from measurements of Stokes profiles on a regular basis. Based on the photospheric field as an input, a force-free field approach, assuming magnetic forces to dominate, can be used to model the magnetic field configuration in the atmospheric layers above. Those models are characterized by the equations
\begin{eqnarray}
	\nabla \times \Bvec &=& \alpha\,\Bvec \label{eq:force}\\
	\nabla \cdot \Bvec &=& 0 \label{eq:divergence}
\end{eqnarray}
and $\Bvec$\,$=$\,$\Bobs$ on the photosphere. Here, $\alpha$ is a scalar torsion parameter, being constant along field lines because of Equation (\ref{eq:divergence}). Since, by assumption, the magnetic energy density is dominating in the corona, force-free approximations are a valid representation of the coronal magnetic field (except for the very moment of explosive events).

Simpler force-free approaches (potential and linear force-free, assuming $\alpha\,$=$\,0$ and $\alpha$\,$=$\,$constant$, respectively) can describe the coronal magnetic field topology only to some extent since they adhere to restrictive assumptions about the electric current. Such models require only the line-of-sight component of the photospheric magnetic field as an input. Large-scale electric currents, however, are induced by the reconfigurations of the transverse magnetic field component in which excess energy is stored \citep{pri_for_02}. Consequently, potential and linear force-free models can not approximate the magnetic energy content of the corona adequately and nonlinear force-free (NLFF) models are required. NLFF models assume $\alpha$\,$=$\,$\alpha(\rvec)$, \ie, the torsion parameter to be a function of space and varying from field line to field line. These are currently the most realistic class of force-free fields used to model the coronal magnetic field. In particular, it is the excess of magnetic energy of a NLFF field over the corresponding linear force-free field (or a fraction of it), representing an upper limit of the ``free'' magnetic energy that can be either transformed or released during eruptions.

Several studies have focused on the NLFF field reconstruction of the coronal magnetic field and its associated magnetic energy above solar active regions prior to and after flares. Smaller flares were found to require free energies $\propto$\,$10^{23}\J$ -- $10^{24}\J$ \citep{gil_whe_11,jin_tan_10,tha_wie_08b,reg_pri_07b,reg_ama_02} while large flares seem to occur only when energies $\propto$\,$10^{25}\J$ -- $10^{26}\J$ are available \citep{sun_hoe_12,jin_tan_10,jin_che_09,jin_wie_08,tha_wie_08a,schr_08,met_lek_05}.

Force-free field models are applicable for magnetically dominated atmospheric layers which, following \cite{gar_01}, is true between about 400 km and 200 Mm above photospheric levels. Thus, vector maps derived from routine measurements of the Stokes profiles at photospheric levels, represent an inconsistent lower boundary condition for the NLFF extrapolation. ``Preprocessing'' of photospheric magnetic field vector data is a numerical procedure that is advantageous in order to obtain suitable lower boundary conditions for successful force-free reconstructions \citep{fuh_see_11,met_08,wie_tha_08,fuh_see_07}. Moreover, uncertainties in the measurements, caused, \eg, by noise or the particular inversion and azimuth ambiguity removal techniques, should be taken into account \citep{der_schr_09}. The effect of measurement and analysis errors on the outcome of NLFF field extrapolations was studied by \citet{wie_yel_10}. They found that mainly the transverse photospheric field is influenced by instrumental effects and that these errors lead to strongest deviations in resulting NLFF fields at low heights in the model volume.

In this work, we use SDO/HMI and SOLIS/VSM photospheric magnetic field measurements to model the NLFF coronal field above a solar active region. We compare quantities like the total magnetic energy content and free magnetic energy, the longitudinal distribution of the magnetic pressure, and orientation of the magnetic field components in the HMI- and VSM-based model volumes. We relate the appearing differences to the photospheric quantities such as the magnetic fluxes and electric currents but also show the extent of agreement of NLFF extrapolations from different data sources.

\section{Data Sources and Event Selection}

\subsection{Solar Dynamics Observatory - Helioseismic and Magnetic Imager}

The Helioseismic and Magnetic Imager (HMI) is part of the Solar Dynamics Observatory (SDO; \citet{scho_sche_12}) and observes the full Sun at six wavelengths and full Stokes profile in the Fe~{\sc{i}} 617.3 nm spectral line. Photospheric line-of-sight LOS and vector magnetograms are retrieved from filtergrams with a plate scale of 0.5 arc-second (Hoeksema T. \etal, 2012, in preparation). From filtergrams averaged over about ten minutes, Stokes parameters are derived and inverted using the Milne-Eddington (ME) inversion algorithm of \citet{bor_tom_10} (the filling factor is held at unity). Within automatically identified regions of strong magnetic fluxes \citep{tur_jon_10}, the 180$\degree$ azimuth ambiguity is resolved using the Minimum Energy Algorithm \citep[MEA;][]{met_94, met_lek_06, lek_bar_09}. The noise level is $\approx$\,$10\G$ and $\approx$\,$100\G$ for the longitudinal and transverse magnetic field, respectively.

\subsection{Synoptic Optical Long-term Investigations of the Sun - Vector-SpectroMagnetograph}

The Vector-SpectroMagnetograph (VSM) is a part of the Synoptic Optical Long-term Investigations of the Sun project (SOLIS; \citet{kel_har_03a}). The VSM provides, among other measurements, full disk vector observations and area scans of selected active regions on a daily basis, observing conditions permitting. For a  typical observing day, full-disk photospheric vector magnetograms are available as a ´´quick-look´´ (QL) and a ME inversion product \citep{hen_06} based on observations in the Fe~{\sc{i}} 603.15 nm and 603.25 nm spectral lines. 

The QL parameters include estimates of the magnetic flux density, inclination and azimuth. Its main purpose is to serve as an initial guess for the ME inversions. The QL analysis estimates the azimuth and inclination following \citet{aue_hea_77}. The magnetic flux density is approximated using the estimated inclination with the estimated LOS flux \citep{ree_sem_79} and assuming that the filling factor is one.

The vector field parameters, for pixels with polarization signals clearly above the noise levels, are determined using the ME inversion technique as implemented by \citet{sku_lit_87} and based on \citet{aue_hea_77b}. The 180$\degree$ azimuth ambiguity in the transverse field is resolved on the entire solar disk using the Non-Potential magnetic Field Calculation method of \citep[NPFC;][]{geo_05}. The vector data are provided with a plate scale of one arc-second. The lower limits for the noise levels are a few Gauss in the longitudinal and $70\G$ in the transverse field measurements.


\subsection{Event Selection}

\begin{figure}
   \vspace{-1.25cm}
   \includegraphics[width=\columnwidth]{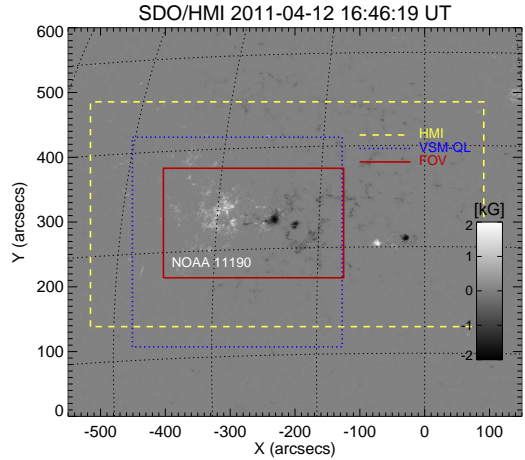}
   \vspace{-1.5cm}
   \caption{HMI full-disk LOS magnetic field on 2011 April 12 at 16:46:19 UT. White/black represents positive/negative magnetic polarities. Yellow dashed outlined is the area for which the 180$\degree$ azimuth ambiguity is resolved for the HMI vector map. The blue dotted outline marks the VSM-QL measurement and the red solid rectangle shows the selected FOV for analysis. The color bar indicates the longitudinal magnetic flux density.}
   \label{fig:fig1}
\end{figure}

To compare the characteristics of a solar active region, full-disk LOS magnetic field observations of the SDO/HMI and the SOLIS/VSM were scanned. {\jt On 2011 April 12, four active regions, AR 11185, AR 11186, AR 11187 and AR11190 were observed on the solar disk.} The latter two, AR 11187 and AR 11190 appear to be well isolated. For AR 11190 both HMI and VSM (QL and ME) data are available (see Figure~\ref{fig:fig1}). Vector magnetic field measurements were available from SDO/HMI at 16:48 UT and from SOLIS/VSM at 16:43 UT. We define our field-of-view (FOV) as $-404''$\,$\le$\,$x$\,$\le$\,$-125''$ and $214''$\,$\le$\,$y$\,$\le$\,$381''$ relative to Sun center, \ie, covering about $200$\,$\times$\,$120\Mm^2$ (red outline around AR 11190 in Figure~\ref{fig:fig1}). The selected FOV is centered around N15E15, with the easternmost regions being located at distances not farther than about $25\degree$ away from disk center. We assume, for the purpose of this work, that the vertical and longitudinal field are almost identical and that the same is true for the horizontal and transverse field. Furthermore, we neglect the curvature of the Sun and assume a planar geometry of the photospheric boundary.

The purpose of this work is to compare the analysis of the photospheric magnetic field and subsequent force-free modeling based on HMI and VSM (ME and QL) vector maps. We investigate if using the not fully ME inverted VSM-QL data product significantly changes the quality of the extrapolation.

\section{Results}

\subsection{Magnetic structure of the Photospheric Field}\label{sec:phot}

\begin{figure*}
   \epsscale{1.2} 
   \plotone{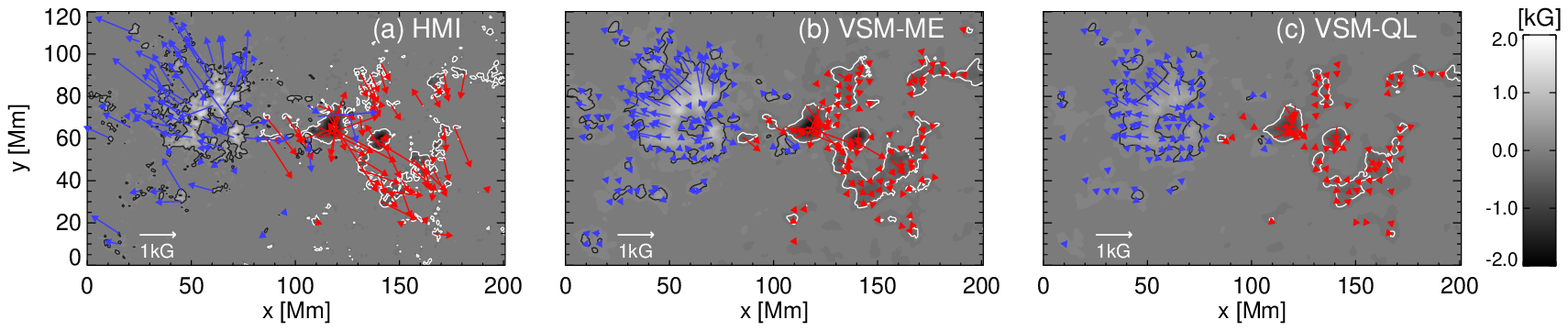}\\
   \plotone{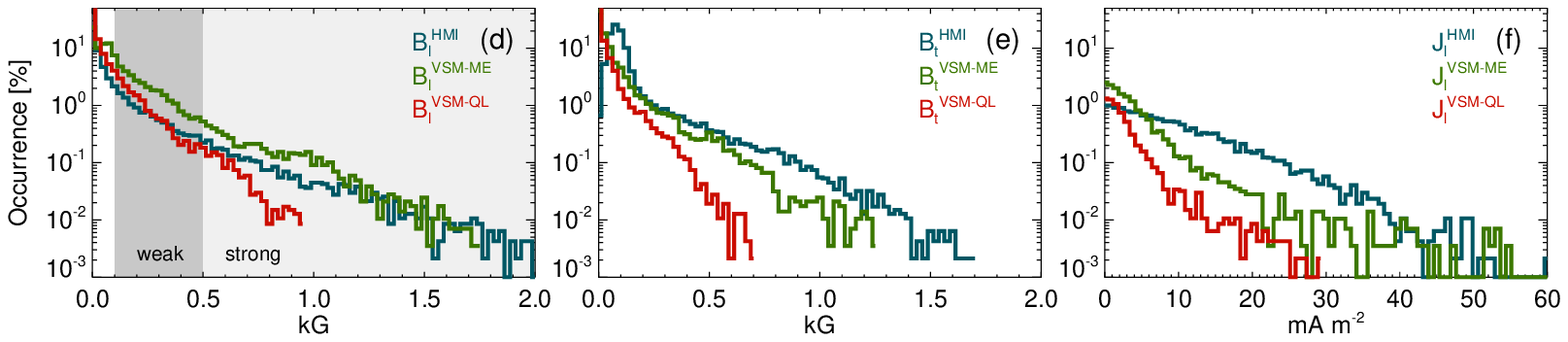}\\
   \caption{Magnetic vector maps from (a) HMI and (b) VSM-ME and (c) VSM-QL. The longitudinal magnetic flux is shown as color-coded background. Blue (red) arrows indicate the transverse magnetic field, originating from (positive) negative polarity for regions where $\Blabs$\,$\ge$\,$100\G$. Contours are drawn at $\pm$\,$200\G$. Histograms of (d) the absolute longitudinal flux $\Blabs$, (e) the absolute transverse field $\Btabs$ and (f) the longitudinal current density $\Jlabs$ (calculated at only where $\Btrans$\,$\ge$\,$100\G$ and $\Blabs$\,$\ne$\,$0$).  HMI, VSM-ME and VSM-QL are represented by blue, green and red lines, respectively. Indicated by the dark and light gray shaded areas in panel (d) are the ``weak'' and ``strong'' field regions for which the longitudinal fluxes are listed in Table \ref{tab:tab1}.}
   \label{fig:fig2}
\end{figure*}

\begin{table}
   \begin{center}
	 \caption{\label{tab:tab1} Unsigned longitudinal magnetic flux $\fu$ calculated from HMI and VSM magnetic vector maps. Indicated are the values for taking all pixels into account (``total"), considering only strong field regions (where $\Blabs$\,$\geq$\,$500\G$; ``strong'') and only weak field regions (where $100\G$\,$\leq$\,$\Blabs$\,$<$\,$500\G$; ``weak''). Also given are the values of the net Lorentz force ($\eforce$) and net torque ($\etorque$).}
	 \begin{tabular}{lccccc}
	    \tableline
	    \tableline            
	    $Data$ & \multicolumn{3}{c}{$\fu$ $[\,\times$\,$10^{22}\Mx\,]$} & $\eforce$ & $\etorque$ \\
	    ~ & $total$ & $strong$ & $weak$ & ~ & ~\\
	    \tableline
	    HMI		& 1.46	& 0.57 & 0.62 & 0.49 & 0.44 \\
	    VSM-ME	& 2.03 & 0.65 & 1.09 & 0.82 & 0.87 \\
	    VSM-QL	& 1.35 	& 0.17 & 0.81 & 0.78 & 0.85 \\
	    \tableline
	    \tableline
	 \end{tabular}
   \end{center}
\end{table}

Given the twice as large plate scale of VSM, we bin the HMI vector maps to the resolution of VSM in order to compare the photospheric magnetic field and subsequent force-free modeling. We note here that this only slightly affects the findings described in the following. The magnetic field in AR 11190 has a dipole structure (Figure~\ref{fig:fig2}a--c). Similar structures are seen in the HMI and VSM data, \eg, the relatively compact appearance of the positive and the more fragmented pattern of the negative polarity region. The transverse field, $\Btrans$, has a clearly diverging (converging) structure above the positive (negative) polarity.

\begin{figure*}
   \epsscale{1.2}
   \plotone{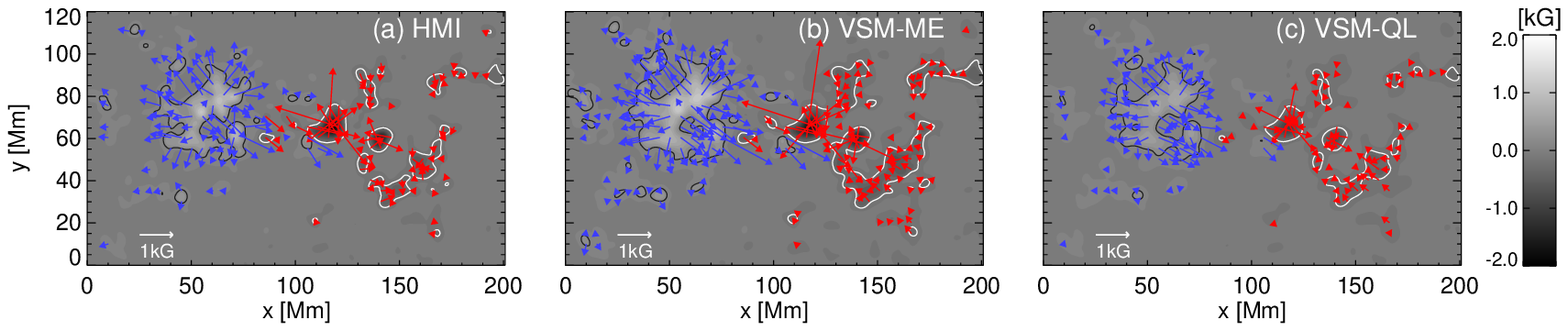}\\
   \plotone{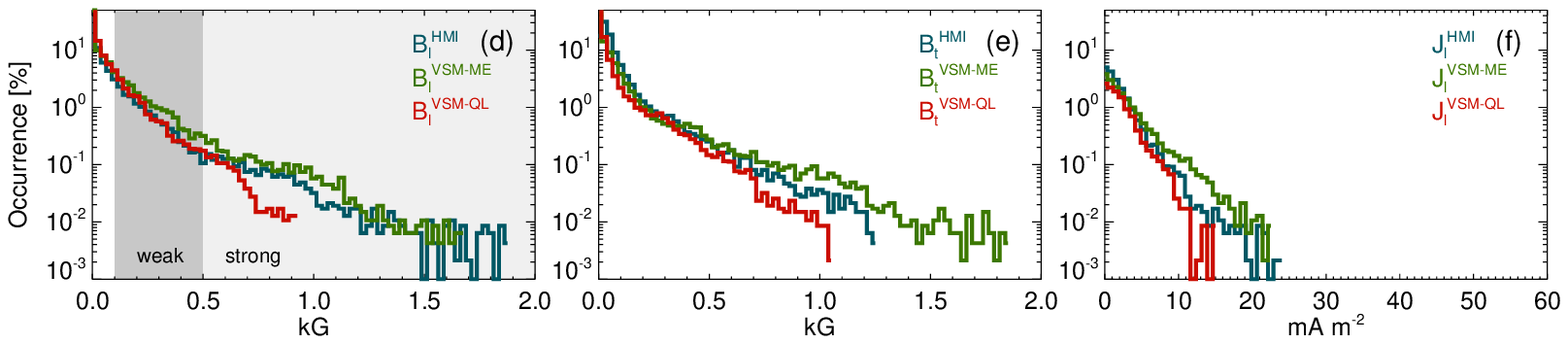}\\
   \caption{Preprocessed magnetic vector maps from (a) HMI and (b) VSM-ME and (c) VSM-QL. The longitudinal magnetic flux is shown as color-coded background. Blue (red) arrows indicate the transverse magnetic field, originating from (positive) negative polarity for regions where $\Blabs$\,$\ge$\,$100\G$. Contours are drawn at $\pm$\,$200\G$. Histograms of (d) the absolute longitudinal flux $\Blabs$, (e) the absolute transverse field $\Btabs$ and (f) the longitudinal current density $\Jlabs$ (calculated at only where $\Btrans$\,$\ge$\,$100\G$ and $\Blabs$\,$\ne$\,$0$). HMI, VSM-ME and VSM-QL are represented by blue, green and red lines, respectively. Indicated by the dark and light gray shaded areas in panel (d) are the ``weak'' and ``strong'' field regions for which the longitudinal fluxes are listed in Table \ref{tab:tab2}.}
   \label{fig:fig3}
\end{figure*}

The unsigned magnetic flux, when calculated from the HMI ($\fuhmi$), VSM-ME ($\fuvsmme$) and VSM-QL ($\fuvsmql$) data is balanced to within $\approx$\,$10\%$. All are on the order of $10^{22}\Mx$, where $\fuhmi$\,$\approx$\,$0.7$\,$\fuvsmme$ and $\fuhmi$\,$\approx$\,$1.1$\,$\fuvsmql$ (see Table \ref{tab:tab1}). The net flux of the HMI and VSM vector maps is positive, $\fsign$\,$\propto$\,$10^{21}\Mx$ and about 10\% of the unsigned flux. In the VSM-ME vector maps, zeros are assigned to pixels where the measured polarization signal is too weak to perform a reliable inversion. If we disregard these ``zero-pixels'', about $60\%$ of the total number of pixels in the HMI and VSM vector maps remain for comparison. Then the unsigned flux in the HMI (VSM-QL) vector map is $\approx$\,$3\%$ less and the net longitudinal flux is $\approx$\,$1\%$ ($\approx$\,$4\%$) less (more). In other words, the unsigned and net longitudinal fluxes of the HMI and VSM-QL vector maps are practically unchanged when disregarding the zero pixels in the VSM-ME vector map and their contribution is negligible.

In the following, we study the spatial distribution of the flux and consider regions of strong and weak magnetic flux separately, defining ``strong'' as where $\Blabs$ greater than $500\G$ and ``weak'' as where $100\G$\,$\leq$\,$\Blabs$\,$<$\,$500\G$ (where $\Blabs$\,$=$\,$|\Blos|$). The zero-pixels do not cover parts of the weak and strong field regions, \ie, they don't influence the calculated quantities. For all vector maps the weak flux regions have more flux when compared to the strong flux regions: $\approx 10\%$ more in HMI, $\approx$\,$40\%$ more in VSM-ME and $\approx$\,$80\%$ more in VSM-QL (the latter being partly a consequence of the cutoff of $\Blabsvsmql$ at $\approx$\,$1\kG$). Weak field regions cover 10\% and 20\% of the FOV of HMI and VSM, respectively. In these regions the VSM data have more longitudinal flux ($\fuvsmme$\,$\approx$\,$1.7$\,$\fuhmi$ and $\fuvsmql$\,$\approx$\,$1.3$\,$\fuhmi$). The strong field regions cover around 3\% of the considered area of the HMI and VSM-ME vector maps, and 1\% of the VSM-QL data. In those regions, the HMI (VSM-QL) data reveal only about 90\% (25\%) of $\fuvsmme$. Compared to HMI, VSM-QL has less longitudinal flux in strong field regions but more in weak field regions. This might partly be attributable to the influence of scattered light. The relative portion of scattered light is higher in strong field regions so that the effect on the measured flux is larger, \ie, measured fluxes are more strongly reduced. {\jt This level of scattered light in the VSM vector maps is less than 10\%, while its importance for HMI vector maps is not quantified yet.}

The distribution of $\Blabs$ decays rapidly with increasing flux density (see Figure~\ref{fig:fig2}d). On overall, the longitudinal field of the HMI and VSM-ME vector maps are comparable, especially at higher field strengths. In general, VSM-ME detects more longitudinal field than HMI. While $\Blabsvsmme$ does not exceed $\approx$\,$1.8\kG$, $\Blabshmi$ can be as high as $\approx$\,$2.1\kG$. A different behavior is found for the VSM-QL data. The HMI and VSM-QL estimates match at field strengths below $\approx$\,$0.7\kG$, the relative occurrence of $\Blabsvsmql$ clearly differs at higher values. A similar trend is found when considering the absolute value of the transverse field, $\Btabs$\,$=$\,$|\Btrans|$, (Figure~\ref{fig:fig2}e). Maximal values of $\Btabsvsmme$ are below $\approx$\,$1.3\kG$ while $\Btabshmi$ can be up to $\approx$\,$1.8\kG$. The shapes of the $\Btabs$ distributions differ with larger discrepancies at higher values of $\Btabs$. {\jt Only at values $0.1\kG$\,$\lesssim$\,$\Btabs$\,$\lesssim$\,$0.8\kG$ the distributions of $\Btabsvsmme$ and $\Btabshmi$ are relatively close to each other. Similar to the trend of the longitudinal field, the transverse field estimates of VSM-QL increasingly differ with increasing $\Btabs$ and, on overall, VSM-QL detects less transverse field.}

\begin{table}
   \begin{center}
	 \caption{\label{tab:tab2} Unsigned longitudinal magnetic flux $\fu$ calculated from the preprocessed HMI and VSM magnetic vector maps. Indicated are the values for taking all pixels into account (``total"), considering only strong field regions (where $\Blabs$\,$\geq$\,$500\G$; ``strong'') and only weak field regions (where $100\G$\,$\leq$\,$\Blabs$\,$<$\,$500\G$; ``weak''). Also given are the values of the net Lorentz force ($\eforce$) and net torque ($\etorque$).}
	 \begin{tabular}{lccccc}
	    \tableline
	    \tableline            
	    $Data$ & \multicolumn{3}{c}{$\fu$ $[\,\times$\,$10^{22}\Mx\,]$} & $\eforce$ & $\etorque$ \\
	    ~ & $total$ & $strong$ & $weak$ & ~ & ~\\
	    \tableline
	    HMI		&  1.42 & 0.41 & 0.71 & 0.01 & 0.01 \\
	    VSM-ME	&  2.03	& 0.62 & 1.11 & 0.01 & 0.02\\
	    VSM-QL	& 1.35	& 0.16	& 0.81 & 0.01 & 0.02\\
	    \tableline
	    \tableline
	 \end{tabular}
   \end{center}
\end{table}

We also show the absolute value of the longitudinal current density, $\Jlabs = |\Jvec_l|$, distribution in Figure~\ref{fig:fig2}f. We calculate the longitudinal current density $\Jl$ everywhere where $\Btrans$\,$\ge$\,$100\G$ and $\Blabs$\,$\ne$\,$0$. The $\Jlabs$ distribution clearly differ over the entire range, most pronounced at $10\mAm$\,$\lesssim$\,$\Jlabs$\,$\lesssim 30\mAm$. On overall, $\Jlabs$ is strongest in the HMI vector maps and weakest in the VSM-QL data. The mean values of $\Jlabs$ are $5\mAm$, $0.6\mAm$ and $0.7\mAm$, respectively. The spectral power of $\Jlabs$ is more or less constant on all spatial scales in the HMI vector maps. For the VSM vector maps more spectral power of $\Jlabs$ is found on larger scales  than on small scales. The total longitudinal current is approximately $-4$\,$\times$\,$10^7\A$, $-2$\,$\times$\,$10^7\A$ and $-1$\,$\times$\,$10^7\A$ for the HMI, VSM-ME and VSM-QL vector maps, respectively.

{\jt We also probe the influence of the different methods used to resolve the 180$\degree$ azimuth ambiguity in the transverse field of  the VSM and HMI vector maps. We compare $\Blabs$, $\Btabs$ and $\Jlabs$ of the MEA ambiguity resolved HMI vector maps to that of NPFC ambiguity resolved HMI maps (courtesy of M. Georgoulis). The power spectra of $\Blabs$, $\Btabs$ or $\Jlabs$ as a function of wave number (\ie, length scale) are very similar and show no significantly different distributions. The power of $\Blabs$ and $\Btabs$ decreases with increasing wave number (\ie, with decreasing length scale, prominently on scales $\lesssim$\,$20\Mm$) while the power of $\Jlabs$ if well distributed over all length scales. We therefore assume that the method used to resolve the 180$\degree$ azimuth ambiguity has a minor impact on $\Blabs$, $\Btabs$ and $\Jlabs$.}

 All of the above indicates that most longitudinal magnetic flux is detected by VSM-ME, while the transverse field, and hence electric currents, are partly of spatial scales which VSM-ME cannot fully resolve. Weakest signals for all, $\Blabs$, $\Btabs$ and $\Jlabs$, is found in the VSM-QL data. However, as will be shown below, the relative occurrences of all quantitites are clearly more similar after applying a ``preprocessing'' routine.

\subsubsection{The Effect of Preprocessing}\label{ss:pp}

Vector maps, suitable as input for our force-free model technique are expected to be flux-balanced with a vanishing  net force and net torque. We use dimensionless parameters, $\eforce$ and $\etorque$, to quantify such properties \citep[]{wie_inh_06,aly_89,mol_74,mol_69}. For AR 11190, $\eforce$ and $\etorque$ are on the order of $1.0$ (see Table \ref{tab:tab1}). These are similar to the values calculated from vector data used in previous studies \citep[Wiegelmann T., \etal, 2012, \solphys, submitted;][]{tha_wie_08b}, showing that the measured data are, in principle, inconsistent with the force-free field approach. We apply the preprocessing routine of \citet{wie_inh_06} which reduces the net force and net torque. Doing this, we do not intend to suppress the existing forces in the photosphere (where we know that the field is not force-free). Instead, we try to approximate the magnetic field at a chromospheric level where magnetic forces are expected to be much smaller than in the layers below. From the preprocessed vector maps, which we assume to be situated somewhere between $400\km$ and $800\km$ above the photosphere (\eg, \citet{gar_01,met_95}), we find $\eforce$\,$\propto$\,$10^{-2}$ and $\etorque$\,$\propto$\,$10^{-2}$ (see Table \ref{tab:tab2}).

The preprocessing influences the structure of the magnetic vector data. As can be seen in Figure~\ref{fig:fig3}a--c, the preprocessed $\Btrans$ of HMI, VSM-ME and VSM-QL are more similar than before applying the preprocessing (as shown in Figure~\ref{fig:fig2}a--c). This is because the preprocessing scheme, not only smooths $\Btrans$ but also alters its values in order to reduce the net force and torque. This is physically motivated since $\Btrans$ is measured with lower accuracy than the longitudinal magnetic field. The effect of smoothing can be recognized by comparing the $\pm$\,$0.2\kG$ contour of the original and preprocessed HMI vector map (Figure~\ref{fig:fig2}a and Figure~\ref{fig:fig3}a, respectively). It is less fragmented in the preprocessed data, compared to that in the original data.

As in the original data, the preprocessed HMI, VSM-ME and VSM-QL data have more longitudinal flux in the weak field regions than in the strong fields ($\approx$\,$30\%$, $\approx$\,$40\%$ and $\approx$\,$80\%$ more, respectively; see Table \ref{tab:tab2}). As for the raw data, the strong field regions cover about 2\%, 3\% and 1\% of the considered area of HMI, VSM-ME and VSM-QL. In those regions, the HMI data reveal $\approx$\,$80\%$ and VSM-QL $\approx$\,$30\%$ of $\fuvsmme$. Weak field regions cover about 15\% and 20\% of the FOV of HMI and VSM, respectively. In these regions, the VSM-ME data have most longitudinal flux ($\fuvsmme$\,$\approx$\,$1.6$\,$\fuhmi$), followed by the VSM-QL data ($\fuvsmql$\,$\approx$\,$1.2$\,$\fuhmi$).

The relative occurrences of $\Blabs$ in the preprocessed HMI, VSM-ME and VSM-QL vector maps are clearly more consistent with each other than for the raw data (Figure~\ref{fig:fig3}d). This is to be expected since the preprocessing scheme only smooths the longitudinal field (\ie, does not account for any flux imbalance) while it smooths and alters the transverse field. The latter obviously results in the histograms of $\Btabs$ agreeing well over the entire range of strengths after the preprocessing (Figure~\ref{fig:fig3}e). We suspect that this is due to smoothing of the instrumental noise. {\jt The mean value of the changes due to preprocessing in the longitudinal field is $\propto$\,$10^{-7}\G$ and for the transverse field on the order of $10\G$ (about 1\% of the peak transverse flux). Therefore, the mean changes due to the preprocessing are well within the measurement uncertainty of the HMI and VSM (though the largest local modifications can be, for both $\Blabs$ and $\Btabs$, several $100\G$).} The modification due to preprocessing also includes the rotation of $\Btrans$ (\ie, both $\Bx$ and $\By$ change sign). This affects 2.0\%, 3.2\% and 0.8\% of the locations where $\Btabs$\,$>$\,$250\G$ in the HMI, VSM-ME and VSM-QL vector maps, respectively. For pixels with transverse fields on the order of the measurement error or less, the field is rotated in 6.9\% of the cases in the HMI and VSM-ME vector maps and in 3.3\% of the cases in the VSM-QL data. By coincidence, this rotation could cause a flipping of the transverse field (\ie, a rotation of 180$\degree$) but this was not observed when comparing the original with the preprocessed data.

Altering the transverse magnetic field components by preprocessing necessarily leads to a change in the electric currents. The relative occurrence of $\Jlabs$ (Figure~\ref{fig:fig3}f) shows that the currents of the preprocessed HMI and VSM-ME vector maps are very similar. Comparing, Figures \ref{fig:fig2}d and \ref{fig:fig3}d, it can be seen that the highest absolute values $\Jlabs$ decrease, about $\approx$\,$11\%$ for HMI and about $\approx$\,$18\%$ for the VSM-ME. The mean values of $\Jlabs$ are $\approx$\,$0.9\mAm$, $\approx$\,$0.6\mAm$ and $\approx$\,$0.4\mAm$, respectively. Though reducing the total longitudinal current, the preprocessing does not redistribute the current to larger scales, it causes a decrease of the spectral power of $\Jlabs$ at small scales (especially on scales below $\approx$\,$6\Mm$). The total longitudinal current decreases to about $-1$\,$\times$\,$10^7\A$, $-3$\,$\times$\,$10^6\A$ and $-1$\,$\times$\,$10^6\A$ in the preprocessed HMI, VSM-ME and VSM-QL vector maps, respectively.

\subsection{Force-free Modelling}

As pointed out by \citet{der_schr_09}, even after preprocessing the photospheric vector maps, force-free model techniques might still not find a strictly force- and divergence-free solution in the coronal volume above. Therefore, we calculate the NLFF fields using an extended optimization scheme \citep{wie_inh_10}. {\jt By construction, the NLFF field drops towards a potential field, prescribed on the lateral and top boundaries of the volume-bounding surface, $\volel$, within a finite-size layer during the iterative solution process.} For further computations and subsequent analysis, we extract the inner physical domain (that is the sub-volume, excluding the boundary layer towards the side and top boundaries). The resulting domain of analysis is of the dimension $\volsub$\,$\approx$\,$179$\,$\times$\,$98$\,$\times$\,$75\Mm^3$. We use the normal component of the magnetic field, $\Bn$, on the volume-bounding surface of this NLFF sub-field, $\volels$, to calculate the corresponding potential field $\Bpot$\,$=$\,$\nabla \psi$, where $\psi$ is the solution to the Laplace equation, subject to $\partial_n$\,$\psi$\,$=$\,$\Bnnlff$ on $\volels$. To quantify the quality of the reconstructed NLFF fields, we list the normalized mean Lorentz force ($\Lforce$), the mean field divergence ($\Ldiv$) and the current-weighted mean angle, $\Tj$, between $\Bvec$ and $\Jvec$ (see Appendix \ref{ap:quality}) in Table \ref{tab:mergs}.

The improved optimization scheme allows us to relax the magnetic field also on the lower boundary (besides minimizing the volume-integrated force and divergence; see Appendix \ref{ap:modeling}). The relaxation of the lower boundary introduces a further modification of the vector data, in addition to that by the preprocessing applied before. {\jt The mean value of the change of the longitudinal fields due to the relaxation of the lower boundary is $\propto$\,$10^{-4}\G$ and largest local modifications are of the order $1\G$. The mean changes of the transverse field are on the order of $10 \G$ and largest local modifications can be several $100 \G$. Given the noise levels of HMI and VSM measurements of the longitudinal ($\approx$\,$10 \G$ and a few$\G$, respectively) and transverse field ($\approx$\,$100 \G$ and $\gtrsim$\,$70 \G$, respectively), the mean (largest local) modifications are well within (on the order of) the measurement error.} In addition to the modification of the absolute fluxes, the transverse field vector is rotated in some places (\ie, the sign of both $\Bx$ and $\By$ changes). For transverse fields stronger than $250\G$, such a rotation occurs in only about 0.1\% of the corresponding pixels in the preprocessed VSM data but nowhere in the preprocessed HMI vector map. The rotation predominately occurs in regions of weak transverse fields close to the measurement uncertainties (at 17.7\%, 14.7\% and 1.7\% of the pixels with $\Btabs$\,$\le$\,$100\G$ in the preprocessed HMI, VSM-ME and VSM-QL data, respectively). Comparing the preprocessed vector maps and the lower boundary after optimization, no places were found where a flipping (\ie, a rotation by 180$\degree$) occurred.

As can be seen in Table \ref{tab:mergs} (see also Appendix \ref{ap:calcquants}), the energy of the potential and the NLFF field, extrapolated from the VSM-ME data is a factor of $\approx$\,$2$ higher than that of the model based on HMI or VSM-QL data. This is also true for the upper limit for the free magnetic energy, $\Efree$. The unsigned longitudinal flux on the VSM-ME bottom boundary has $\approx$\,$30\%$ more unsigned longitudinal flux than the HMI and VSM-QL lower boundary. Since the longitudinal field component enters in form of boundary conditions in the potential field calculation, it is not surprising that the VSM-ME field model has more potential energy. The energy of the NLFF field is expected to be higher than the potential energy, since it also incorporates the transverse field components. This is true for all models. The relative amount of energy, available to be set free during reconfigurations of the magnetic field, $\Efree$, is $\approx$\,$7\%$, $\approx$\,$11\%$ and $\approx$\,$14\%$ of the respective NLFF field energy, $\Enlff$, for the HMI, VSM-ME and VSM-QL models. Based on the preprocessed lower boundaries, we also calculate the total force-free energy given by the magnetic virial theorem (see \citet{met_lek_05} and Equation \ref{eq:evir} in Appendix \ref{ap:calcquants}). We find a virial energy $\Evir$\,$\propto$\,$10^{25}\J$ ($\Evirvsmme$\,$>$\,$\Evirhmi$\,$>$\,$\Evirvsmql$) and comparable to $\Enlff$. In particular, we find that for the HMI (VSM) models, the virial energy is higher (lower) than the NLFF field energy.

\begin{table}
   \begin{center}
	 \caption{\label{tab:mergs} Virial ($\Evir$) and total magnetic energy ($\Enlff$) of the NLFF field, potential field energy ($\Epot$) and the upper limit for the free magnetic energy ($\Efree$\,$=$\,$\Enlff$\,$-$\,$\Epot$). Normalized mean Lorentz force ($\Lforce$), mean field divergence ($\Ldiv$) and current-weighted mean angle, $\Tj$.}
	 \begin{tabular}{l|cccc|cc|c}
	    \tableline
	    \tableline            
   	    $Data$ & $\Evir$ & $\Enlff$ & $\Epot$ & $\Efree$ & $\Lforce$ & $\Ldiv$ & $\Tj$\\
	    ~ & \multicolumn{4}{|c|}{$[\,\times$\,$10^{25}~{\rm J}\,]$} & ~ & ~ & $[\,\degree\,]$\\
	    \tableline
	    HMI 	& 2.97 & 2.68 & 2.49 & 0.19 & 0.43 & 0.24 & 3.8 \\
	    VSM-ME	& 4.12	& 5.14 	& 4.55	& 0.59 & 0.54	& 0.33	& 5.2\\ 
	    VSM-QL	& 1.89	& 2.14	& 1.85	& 0.29 & 0.50	& 0.33	& 4.4\\ 
	    \tableline
	    \tableline
	 \end{tabular}
   \end{center}
\end{table}

\begin{figure*}
   \epsscale{1.2} 
   \plotone{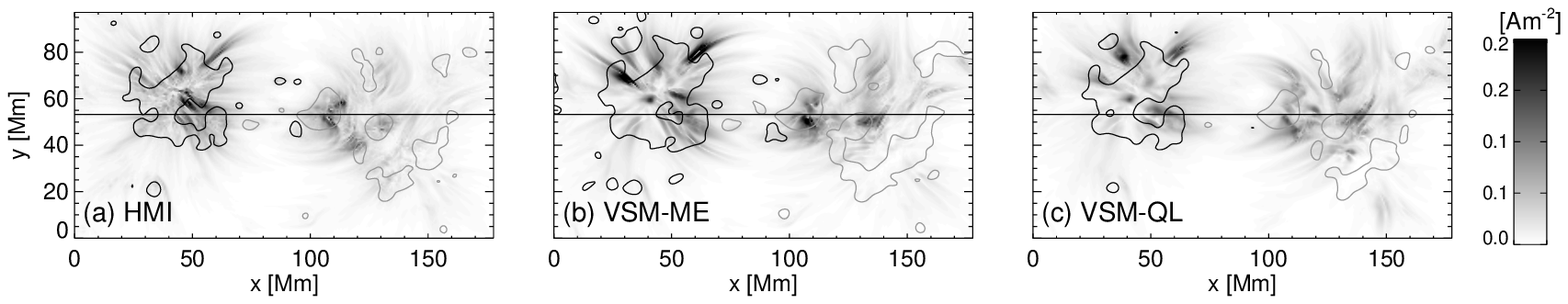}\\
   \plotone{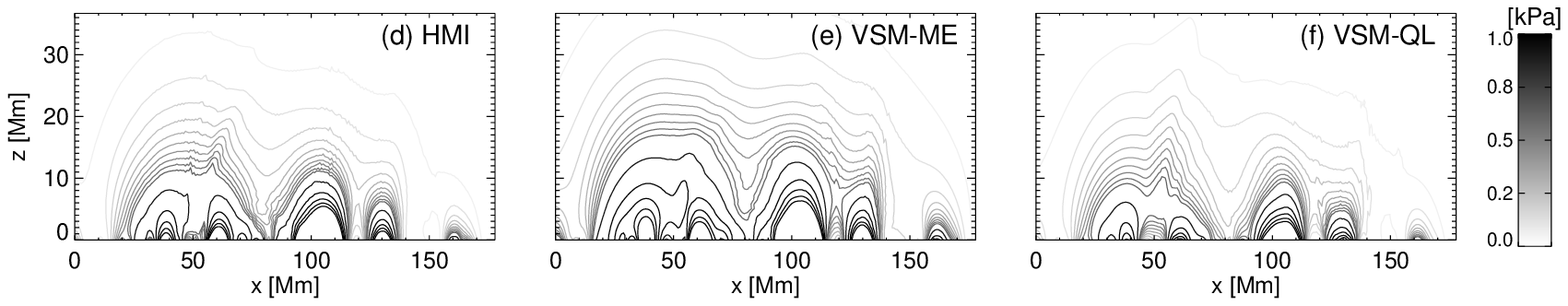}\\
   \plotone{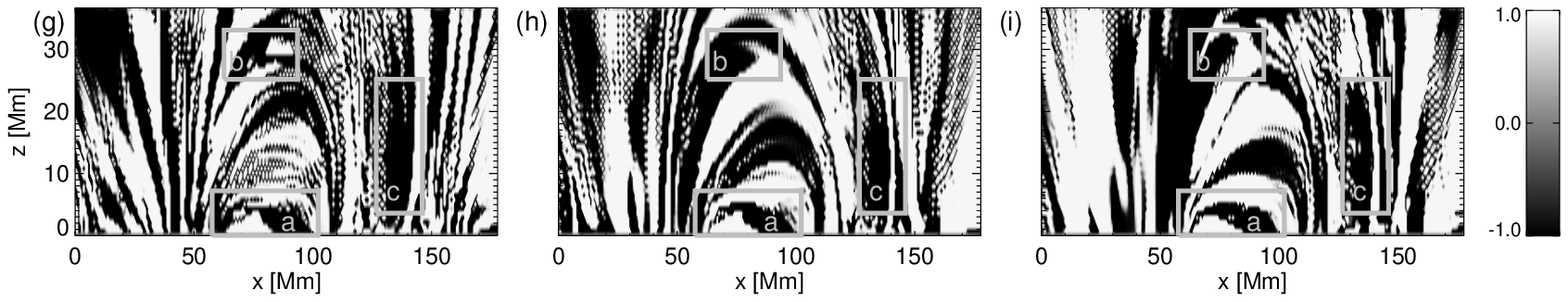}
   \caption{Vertically integrated current density (where $\Btrans$\,$\ge$\,$100\G$), computed from the NLFF extrapolated fields of (a) HMI, (b) VSM-ME and (c) VSM-QL, over the lowest $36\Mm$. Black (gray) contours of the longitudinal field at the bottom of the model volume are drawn at +($-$)$200\G$. Magnetic pressure $\pmag$ below $36\Mm$ in the longitudinal cross-section at $y$\,$=$\,$53\Mm$ for the (d) HMI, (e) VSM-ME and (f) VSM-QL NLFF models. Panels (g), (h) and (i) show the cosine of the angle between the magnetic field and the electric current in the same longitudinal cross-section. The footprint of the longitudinal cross-section on the lower boundary is indicated by the black horizontal line in Figure~\ref{fig:fig4}a-c.}
   \label{fig:fig4}
\end{figure*}

The absolute value of the longitudinal electric current densities, $\Jlabs$, integrated over the lowest $36\Mm$ in the NLFF field models based on HMI, VSM-ME and VSM-QL data is shown in Figure~\ref{fig:fig4}a--c. (We calculate the current density only at locations in the volume where $\Btrans$\,$\ge$\,$100\G$ and $\Blabs$\,$\ne$\,$0$.) Their origin in the underlying positive ($20\Mm$\,$\lesssim$\,$x$\,$\lesssim$\,$70\Mm$) and negative ($90\Mm$\,$\lesssim$\,$x$\,$\lesssim$\,$130\Mm$) magnetic polarity regions is clearly visible. The HMI-based integrated $\Jlabs$ shows more fine structures than that based on VSM data. Above the positive polarity area we find a, more or less, radially outward directed current structure. Above the negative polarity region, the current-related structures are more diffusely distributed which might be due to the underlying photospheric field being more fragmented. There are also strong, isolated current patches in the VSM-ME model (\eg, in the north-eastern part of the AR, above the positive polarity region) which are weaker in the HMI and VSM-QL models. On overall, the pattern of the longitudinally integrated current density is the same for the HMI and VSM model fields.

\begin{figure*}
   \epsscale{1.2}
   \plotone{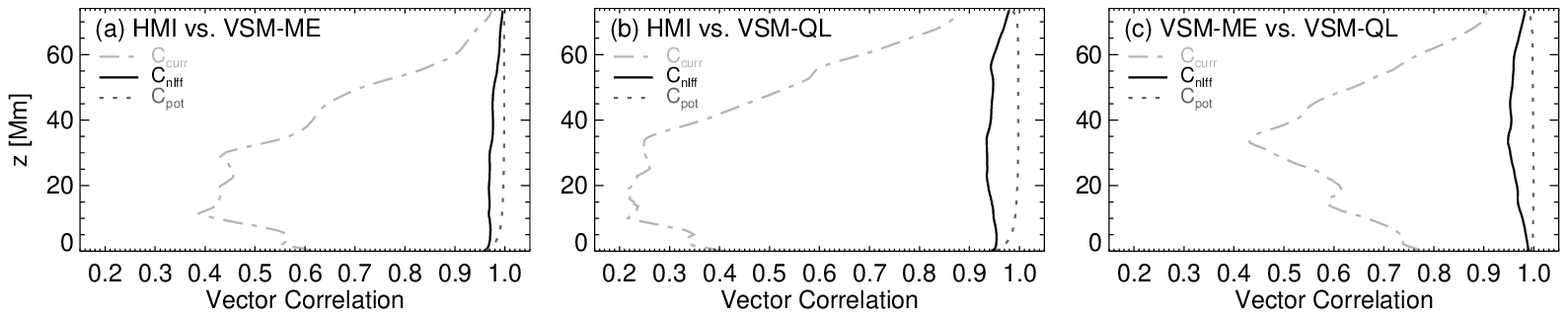}
   \caption{Vector correlation of the model fields as a function of height in the volume of interest. The correlation between the potential fields, $\Cpot$, is shown as dark gray dotted lines. The correlation of the NLFF fields, $\Cnlff$, is drawn as black solid lines. {\jt That of the current-carrying part of the fields ($\Ccurr$; with $\Bcurr$\,$=$\,$\Bnlff - \Bpot$) is represented by the light gray dashed-dotted lines.} The comparison HMI and VSM-ME models is shown in (a), that of the HMI and VSM-QL models in (b), and that of the VSM-ME and VSM-QL models is displayed in (c).}
   \label{fig:fig5}
\end{figure*}

We also investigate the magnetic pressure, $\pmag$, in a longitudinal cross-section at $y$\,$=$\,$53\Mm$ (as indicated by a horizontal line in Figure~\ref{fig:fig4}a-c) of the NLFF fields (Figure~\ref{fig:fig4}d-f). The overall pattern of $\pmag$ appears to be the same when calculated from the HMI ($\pmaghmi$), VSM-ME ($\pmagvsmme$) and VSM-QL ($\pmagvsmql$) NLFF model volume. Again, correspondence to the underlying positive and negative polarity regions is visible. On overall, $\pmagvsmme$\,$>$\,$\pmaghmi$\,$>$\,$\pmagvsmql$ for same locations in the cross section. This is expected, since $\pmag$ is proportional to $\Em$, for which a similar trend was found. Iso-contours, indicating the magnitude of $\pmagvsmme$, are located higher up in the atmosphere compared to $\pmaghmi$ and $\pmagvsmql$, indicating that the VSM-ME NLFF field is sustained by a higher magnetic pressure from below. Three sub-systems are clearly identifiable above the positive polarity region, around $x$\,$\approx$\,$50\Mm$, most pronounced in $\pmaghmi$ and subsequently more diffuse in $\pmagvsmql$ and $\pmagvsmme$. A longitudinal dip-like structure is clearly visible in $\pmagvsmme$ and $\pmagvsmql$ at altitudes $z$\,$\gtrsim$\,$10\Mm$ but very weak in $\pmaghmi$. Above the negative polarity region, another two-component sub-system is revealed, above $105\Mm$\,$\lesssim$\,$x$\,$\lesssim$\,$140\Mm$.

To visualize the alignment of the magnetic field, $\Bvec$, and the electric current, $\Jvec$, we show $\cosxi$\,$=$\,$(\Jvec \cdot \Bvec)/(\Babs\,\Jabs)$ in the longitudinal cross-section at $y$\,$=$\,$53\Mm$ (Figure~\ref{fig:fig4}g--i). A value of zero means that the current is perpendicular to the magnetic field. A value of ($-$)+1 indicates that the current is (anti)parallel to the magnetic field, as it should be for a perfectly force-free field. On overall the same basic alignment pattern of $\cosxi$ is found in all model fields though they are shifted and deformed when compared to each other. Some distinct features are clearly recovered, \eg, a low-lying arc-like structure (region ``a'' in Figure~\ref{fig:fig4}g--i). Similarly, a dip-like structure as well as a vertically oriented pattern are discernible in all models (regions ``b'' and ``c'' in Figure~\ref{fig:fig4}g--i, respectively). However, there are also places where the electric current is parallel to the HMI field while it is anti-parallel to the VSM-ME and/or VSM-QL field. Those places cannot be attributed to locations of particularly weak currents and we suppose that these are consequences of the differing non-potential (current-carrying) contributions to the NLFF field solutions (see below). The appearing discrepancies may partly be caused by not looking at exactly the same topological structure in the different models when displaying at a certain longitudinal cross-section of the computational volume.

{\jt We investigate the similarity of the VSM-ME, VSM-QL and HMI models by comparing the vector correlation (see Appendix \ref{ap:calcquants}) of the potential fields ($\Cpot$), the NLFF fields ($\Cnlff$) as well as of the non-potential part (current-carrying; $\Bcurr$\,$=$\,$\Bnlff$\,$-$\,$\Bpot$), $\Ccurr$.} In all cases, $\Cpot$\,$>$\,$ \Cnlff$\,$>$\,$\Ccurr$ with $\Cpot$\,$\simeq$\,$1$ (\ie, the potential fields are almost identical) and $\Cnlff$\,$\gtrsim$\,$0.95$ (see Figure~\ref{fig:fig5}). The average vector correlation between the potential fields based on the HMI and VSM-ME  data as well as between the HMI and VSM-QL data  is 0.98 (dotted lines in Figure~\ref{fig:fig5}a and \ref{fig:fig5}b, respectively). The potential fields based on the VSM-ME and VSM-QL data agree best with $\Cpot$\,$\simeq$\,$1$ (dotted line in Figure~\ref{fig:fig5}c). The average vector correlation between the NLFF fields is 0.97, 0.94 and 0.98 for HMI vs.\,VSM-ME, HMI vs.\,VSM-QL and VSM-ME vs.\,VSM-QL, respectively. The high values of $\Cpot$ and $\Cnlff$ are not surprising since the NLFF fields do not deviate too much from the corresponding potential fields. This can be seen from the relative amount of free magnetic energy of only $\approx$\,$10\%$ for all models. The vector correlation of the current-carrying parts only, however, shows a different behavior. On average, the vector correlation between the current-carrying parts is 0.53, 0.32 and 0.66 for HMI vs.\,VSM-ME, HMI vs.\,VSM-QL and VSM-ME vs.\,VSM-QL, respectively. A varying pattern with height is found, with a poor correlation at low altitudes (below $\approx$\,$40\Mm$) and steadily increasing for higher altitudes. The increasing values with increasing altitude reflect the nearness of the NLFF fields to the potential field solutions. With increasing altitude, the correlation of the potential field solutions is more and more dominant since, by construction, a NLFF field drops to the potential field towards the top boundary. To summarize, though the potential and NLFF field models based on the HMI, VSM-ME and VSM-QL data are similar and show a high vector correlation, this is not true for the current-carrying part of the fields where strong deviations, especially at low altitudes, are found.

\section{Conclusion}

We use photospheric vector maps, observed on 2011 April 12 to model the nonlinear force-free field above AR 11190. We use the ME and QL products from SOLIS/VSM as well as SDO/HMI data. We compare the measured photospheric VSM-ME, VSM-QL and HMI data. We find that VSM-ME detects most longitudinal flux, followed by HMI and VSM-QL. In strong and weak field regions VSM-ME detects more longitudinal flux than HMI. This is not true for VSM-QL: it shows less flux than HMI in weak field regions but more flux in strong field regions. HMI is found to detect most transverse field. The transverse fields are not detected as well in the VSM data. On overall, the longitudinal electric current is strongest in the HMI data, followed by that of VSM-ME and VSM-QL.

Applying a preprocessing routine to construct boundary conditions, compatible with the force-free approach, results in the overall structure of the transverse field of the HMI and VSM vector maps becoming more similar. The longitudinal and transverse magnetic flux of the preprocessed VSM-ME and HMI vector maps agree very well as does the longitudinal electric current. Only the VSM-QL data has a clearly lower signal in all photospheric measures. Using the preprocessed HMI and VSM vector maps as an input, we model the force-free field above AR 11190. Thus, we are able to quantify the effect of the choice of the lower boundary condition on the model outcome. 

The different longitudinal magnetic flux translates to differences in the force-free field energies of the modeled  coronal volumes. The NLFF, potential and, consequently, the free magnetic energies (all $\propto$\,$10^{25}\J$) of the VSM-ME model are a factor of $\approx$\,$2$ higher than those of the HMI and VSM-QL models. The models' free energy is between $\approx$\,$7\%$ and $\approx$\,$14\%$ of the NLFF field energy, corresponding to $\approx$\,$2$ -- $3$\,$\times$\,$10^{24}\J$. The virial energy is comparable to the NLFF field energy and slightly (higher) lower in the (HMI-) VSM-based models.

It is believed that flare sizes relate to the available amount of free energy in the coronal volume. Only one larger event (a M1.3 flare on April 15,  peaking at 17:12 UT) originated from AR 11190. A smaller C 1.1 flare took place on April 13 and a C1.6 flare on April 14. From our force-free models, we find a free magnetic energy $\propto$\,$10^{24}\J$ on April 12. This is certainly enough to power C-class flaring (\eg, \cite{jin_tan_10,tha_wie_08b}). The question arises if this amount of energy plus the additional built up energy during the next two days were sufficient to power the M-class flare about three days later. It cannot be ruled out that this amount of pre-flare energy is sufficient to power the event but it seems rather low.

Agreements of the HMI- and VSM-based extrapolations are also found, \eg, in the vertical distribution of the magnetic pressure and the longitudinally integrated current density. The longitudinal and transverse vector components of the reconstructed NLFF fields match each other to quite high degree when viewed as function of height in the atmosphere. The potential and NLFF field models based on the HMI, VSM-ME and VSM-QL data are similar and show a high vector correlation. This is not true for the current-carrying (non-potential) part of the model fields where strong deviations, especially at low altitudes, are found. The orientation of the electric current density with respect to the magnetic field in a vertical cross-section of the extrapolation volume is found to be similar in portions of the model volumes. However, also discrepancies are found for some places where, \eg, the current is parallel to the field in the model based on the HMI data but anti-parallel in the models based on VSM data. On overall, the co-alignment pattern of the electric current and magnetic field in the different models is distorted which may be attributed to the differing non-potential portions of the NLFF model volumes.

We conclude that NLFF field extrapolations based on HMI, VSM-ME, VSM-QL vector maps result in comparable coronal field models and related quantities. Most importantly, the models agree well regarding the estimated relative free magnetic energy which can be released during explosive events. This allows us to assume that our method to reconstruct the coronal magnetic field is robust to deficiencies in the lower boundary data.

\acknowledgments
SOLIS data are produced cooperatively by NSF/NSO and NASA/LWS. SDO data are courtesy of the NASA/SDO HMI science team. J.\,K.\,T.~acknowledges supported by DFG grant WI 3211/2-1, X.\,S.~is financed by NASA contract NAS5-02139 and T.\,W.~is funded by DLR grant 50 OC 0904. We thankfully acknowledge M. Georgoulis for ongoing collaboration and helpful discussions, and the careful consideration of the referee to improve the manuscript.

\appendix

\section{Nonlinear Force-free magnetic field modeling}\label{ap:modeling}

The improved optimization scheme of \citet{wie_inh_10}, based on a proposed principle by \citet{whe_00} and originally implemented and extended by \citet{wie_04} solves the force-free equations by minimizing
\begin{equation}
	\label{eq:lopt}
   \Lopt = \intv \left(\,\wforce\,B^{-2}\,|(\,\nabla \times \Bvec\,) \times \Bvec|^2 + \wdiv |\nabla \cdot \Bvec|^2\,\right) \dvol + \nu \ints (\,\Bvec - \Bobs\,) \cdot \Wvec \cdot (\,\Bvec - \Bobs\,) \dsurf.
\end{equation}
where $B$\,$=$\,$|\Bvec|$ and $\wforce$ and $\wdiv$ are position-dependent weighting functions. $\wforce$ and $\wdiv$ are set to $1$ in the volume of interest and drop to zero in form of a $\cos$-profile within a finite size boundary layer towards the lateral and top boundaries. This reduces the influence of the boundaries of the computational cube on the solution inside. The boundary layer spans 16 pixels in the extrapolation volumes with the HMI vector maps being binned to the resolution of VSM.

The major improvement of our optimization method is the introducing of the surface term in (\ref{eq:lopt}), \ie, to relax the field also on the bottom boundary. $\Bvec$ and $\Bobs$ denote the iterated and observed magnetic field on the lower boundary of the extrapolation volume. Before relaxation, the observed photospheric field is preprocessed in order to achieve suitable force-free consistent boundary conditions \citep[for details see][]{wie_inh_06}. The free parameters controlling the net force and net torque are set to $\mu_1$\,$=$\,$1$ and $\mu_2$\,$=$\,$1$, respectively. The deviation from the original data is weighted with $\mu_3$\,$=$\,$10^{-3}$ and the level of smoothing is set to $\mu_4$\,$=$\,$10^{-2}$.

The speed with which the lower boundary is injected during the extrapolation is controlled by $\nu$. The elements of the newly introduced space-dependent diagonal error matrix $\Wvec$ are defined inversely proportional to the squared measurement error of the respective field component. Ideally, one would define its elements based on the error introduced by, \eg, the method of inverting the measured Stokes profiles or algorithm used to resolve the $180\degree$ azimuth ambiguity of the transverse field. Such error estimations are, however, not (yet) available so we have to make reasonable assumptions on the nature of $\Wvec$. Therefore, we chose its element controlling the LOS field component, $\wlos$, to be unity and the one controlling the transverse field, $\wtrans$, as small but positive. As in a previous study \citep{wie_tha_12}, we tested different choices of $\Wvec$ and $\nu$. In particular, we tested all combinations of $\wtrans$\,$\propto$\,$[\,\Btabs,\Btabs^2,\sqrt{\Btabs}\,]$ and $\nu$\,$=$\,$[\,10^{-1},10^{-2},10^{-3}\,]$, where $\Btabs$\,$=$\,$|\Btrans|$. We found the NLFF field, calculated based on the choice $\wtrans$\,$\propto$\,$\Btabs^2$ and $\nu$\,$=$\,$10^{-3}$, performs best for both HMI and VSM input data (for the quantification of the performance see Section \ref{ap:quality}).

\subsection{Quality of the Nonlinear Force-free Magnetic Field Modeling}\label{ap:quality}

Three measures to quantify the goodness of the reconstructed fields are used. These are the mean Lorentz force ($\Lforce$) and field divergence ($\Ldiv$) and the current weighted mean angle between the magnetic field and the electric current density ($\Tj$), which are of the form
\begin{eqnarray}
   \Lforce &=& \sum_i \frac{|\Jvec_i \times \Bvec_i|^2}{B_i^2},%
   \label{eq:lforce} \\
   \Ldiv &=& \sum_i |\,\nabla \cdot \Bvec_i\,|^2,%
   \label{eq:ldiv} \\
   \Tj &=& \sin^{-1} \left(\,\sum_i \frac{|\Jvec_i \times \Bvec_i|}{B_i}\,\right) \Big/ \sum_i J_i.%
   \label{eq:lj}
\end{eqnarray}
where $J$\,$=$\,$|\Jvec|$. $\Lforce$ and $\Ldiv$ are normalized to a unit volume. For a perfectly force-free case, all the above quantities vanish. For the extrapolations presented in this study, these values are $\Lforce$\,$=$\,$0.43$, $\Ldiv$\,$=$\,$0.24$ and $\Tj$\,$=$\,$3.8\degree$ for HMI. For the VSM-ME (-QL) NLFF models they are $\Lforce$\,$=$\,$0.54$ ($0.50$), $\Ldiv$\,$=$\,$0.33$ ($0.33$) and $\Tj$\,$=$\,$5.2\degree$ ($4.4\degree$). All metrics are significantly better compared to the corresponding values when not relaxing the lower boundary. For comparison, one then finds $\Lforce$\,$=$\,$23.5$, $\Ldiv$\,$=$\,$12.4$ and $\Tj$\,$=$\,$23.2$ for HMI and $\Lforce$\,$=$\,$26.2$ ($26.3$), $\Ldiv$\,$=$\,$17.8$ ($18.8$) and $\Tj$\,$=$\,$37.3\degree$ ($36.8\degree$) for VSM-ME (-QL).

\subsection{Calculation of Related Quantities}\label{ap:calcquants}

The magnetic energy content of the modeled NLFF fields is calculated as
\begin{equation}
   \Em = \frac{1}{2\,\mu_0}\int_V B^2 \dvol.
\end{equation}
An upper limit for the free magnetic energy can be estimated by subtracting the magnetic energy of a potential field ($\Epot$) from the energy of a NLFF field ($\Enlff$), \ie, $\Efree$\,$=$\,$\Enlff$\,$-$\,$\Epot$.

The magnetic virial theorem can be used to calculate the force-free field energy in a volume solely from its boundary. When applied to the lower boundary, it is given by:
\begin{equation}
   \Evir = \frac{1}{\mu_0}\ints (x\,B_x + y\,B_y) B_z \dsurf.
   \label{eq:evir}
\end{equation}
Strictly speaking, one has to take all the magnetic flux at the lower boundary into account, \ie, no field lines are supposed to connect to the outside of the NLFF field. Furthermore, $\Evir$ depends on the origin of the coordinate system unless the net transverse Lorentz forces vanish \citep{whe_met_06}.

To quantify the degree of agreement between two given vector fields, $\Bvec$ and $\Bvec'$, we use the ``vector correlation'' metric, as introduced by \cite{schr_der_06}. The agreement of the two vector fields is defined as
\begin{equation}
   C(\Bvec,\Bvec') = \frac{\sum_i \Bvec_i \cdot \Bvec'_i}{\sqrt{\sum_i |\Bvec_i|^2 \sum_i |\Bvec'_i|^2}},
   \label{eq:cvec}
\end{equation}
where $\Bvec_i$ and $\Bvec'_i$ are the two arbitrary vectors fields at each point $i$. For two identical vector fields $\Bvec_i$\,$=$\,$\Bvec'_i$, $C(\Bvec,\Bvec')$\,$=$\,$1$.

\bibliographystyle{apj}
\bibliography{Thalmann_etal_bib}

\begin{thebibliography}{43}
\expandafter\ifx\csname natexlab\endcsname\relax\def\natexlab#1{#1}\fi

\bibitem[{{Aly}(1989)}]{aly_89}
{Aly}, J.~J. 1989, \solphys, 120, 19

\bibitem[{{Auer} {et~al.}(1977{\natexlab{a}}){Auer}, {Heasley}, \&
  {House}}]{aue_hea_77}
{Auer}, L.~H., {Heasley}, J.~N., \& {House}, L.~L. 1977{\natexlab{a}}, \apj,
  216, 531

\bibitem[{{Auer} {et~al.}(1977{\natexlab{b}}){Auer}, {House}, \&
  {Heasley}}]{aue_hea_77b}
{Auer}, L.~H., {House}, L.~L., \& {Heasley}, J.~N. 1977{\natexlab{b}},
  \solphys, 55, 47

\bibitem[{{Borrero} {et~al.}(2010){Borrero}, {Tomczyk}, {Kubo},
  {Socas-Navarro}, {Schou}, {Couvidat}, \& {Bogart}}]{bor_tom_10}
{Borrero}, J.~M., {Tomczyk}, S., {Kubo}, M., {et~al.} 2010, \solphys, 273, 267

\bibitem[{{DeRosa} {et~al.}(2009){DeRosa}, {Schrijver}, {Barnes}, {Leka},
  {Lites}, {Aschwanden}, {Amari}, {Canou}, {McTiernan}, {R{\'e}gnier},
  {Thalmann}, {Valori}, {Wheatland}, {Wiegelmann}, {Cheung}, {Conlon},
  {Fuhrmann}, {Inhester}, \& {Tadesse}}]{der_schr_09}
{DeRosa}, M.~L., {Schrijver}, C.~J., {Barnes}, G., {et~al.} 2009, \apj, 696,
  1780

\bibitem[{{Fuhrmann} {et~al.}(2007){Fuhrmann}, {Seehafer}, \&
  {Valori}}]{fuh_see_07}
{Fuhrmann}, M., {Seehafer}, N., \& {Valori}, G. 2007, \aap, 476, 349

\bibitem[{{Fuhrmann} {et~al.}(2011){Fuhrmann}, {Seehafer}, {Valori}, \&
  {Wiegelmann}}]{fuh_see_11}
{Fuhrmann}, M., {Seehafer}, N., {Valori}, G., \& {Wiegelmann}, T. 2011, \aap,
  526, A70

\bibitem[{{Gary}(2001)}]{gar_01}
{Gary}, G.~A. 2001, \solphys, 203, 71

\bibitem[{{Georgoulis}(2005)}]{geo_05}
{Georgoulis}, M.~K. 2005, \apjl, 629, L69

\bibitem[{{Gilchrist} {et~al.}(2012){Gilchrist}, {Wheatland}, \&
  {Leka}}]{gil_whe_11}
{Gilchrist}, S.~A., {Wheatland}, M.~S., \& {Leka}, K.~D. 2012, \solphys, 276,
  133

\bibitem[{{Henney} {et~al.}(2006){Henney}, {Keller}, \& {Harvey}}]{hen_06}
{Henney}, C.~J., {Keller}, C.~U., \& {Harvey}, J.~W. 2006, in \aspcs, Vol. 358,
  Solar polarization 4, ed. R.~{Casini} \& B.~W. {Lites}, 92--95

\bibitem[{{Jing} {et~al.}(2009){Jing}, {Chen}, {Wiegelmann}, {Xu}, {Park}, \&
  {Wang}}]{jin_che_09}
{Jing}, J., {Chen}, P.~F., {Wiegelmann}, T., {et~al.} 2009, \apj, 696, 84

\bibitem[{{Jing} {et~al.}(2010){Jing}, {Tan}, {Yuan}, {Wang}, {Wiegelmann},
  {Xu}, \& {Wang}}]{jin_tan_10}
{Jing}, J., {Tan}, C., {Yuan}, Y., {et~al.} 2010, \apj, 713, 440

\bibitem[{{Jing} {et~al.}(2008){Jing}, {Wiegelmann}, {Suematsu}, {Kubo}, \&
  {Wang}}]{jin_wie_08}
{Jing}, J., {Wiegelmann}, T., {Suematsu}, Y., {Kubo}, M., \& {Wang}, H. 2008,
  \apjl, 676, L81

\bibitem[{{Keller} {et~al.}(2003){Keller}, {Harvey}, \&
  {Giampapa}}]{kel_har_03a}
{Keller}, C.~U., {Harvey}, J.~W., \& {Giampapa}, M.~S. 2003, in Society of
  Photo-Optical Instrumentation Engineers (SPIE) Conference Series, ed. S.~L.
  {Keil} \& S.~V. {Avakyan}, Vol. 4853, 194

\bibitem[{{Leka} {et~al.}(2009){Leka}, {Barnes}, {Crouch}, {Metcalf}, {Gary},
  {Jing}, \& {Liu}}]{lek_bar_09}
{Leka}, K.~D., {Barnes}, G., {Crouch}, A.~D., {et~al.} 2009, \solphys, 260, 83

\bibitem[{{Metcalf}(1994)}]{met_94}
{Metcalf}, T.~R. 1994, \solphys, 155, 235

\bibitem[{{Metcalf} {et~al.}(1995){Metcalf}, {Jiao}, {McClymont}, {Canfield},
  \& {Uitenbroek}}]{met_95}
{Metcalf}, T.~R., {Jiao}, L., {McClymont}, A.~N., {Canfield}, R.~C., \&
  {Uitenbroek}, H. 1995, \apj, 439, 474

\bibitem[{{Metcalf} {et~al.}(2005){Metcalf}, {Leka}, \& {Mickey}}]{met_lek_05}
{Metcalf}, T.~R., {Leka}, K.~D., \& {Mickey}, D.~L. 2005, \apjl, 623, L53

\bibitem[{{Metcalf} {et~al.}(2006){Metcalf}, {Leka}, {Barnes}, {Lites},
  {Georgoulis}, {Pevtsov}, {Balasubramaniam}, {Gary}, {Jing}, {Li}, {Liu},
  {Wang}, {Abramenko}, {Yurchyshyn}, \& {Moon}}]{met_lek_06}
{Metcalf}, T.~R., {Leka}, K.~D., {Barnes}, G., {et~al.} 2006, \solphys, 237,
  267

\bibitem[{{Metcalf} {et~al.}(2008){Metcalf}, {Derosa}, {Schrijver}, {Barnes},
  {van Ballegooijen}, {Wiegelmann}, {Wheatland}, {Valori}, \&
  {McTtiernan}}]{met_08}
{Metcalf}, T.~R., {Derosa}, M.~L., {Schrijver}, C.~J., {et~al.} 2008, \solphys,
  247, 269

\bibitem[{{Molodenskii}(1969)}]{mol_69}
{Molodenskii}, M.~M. 1969, \sovast, 12, 585

\bibitem[{{Molodensky}(1974)}]{mol_74}
{Molodensky}, M.~M. 1974, \solphys, 39, 393

\bibitem[{{Priest} \& {Forbes}(2002)}]{pri_for_02}
{Priest}, E.~R., \& {Forbes}, T.~G. 2002, \aapr, 10, 313

\bibitem[{{Rees} \& {Semel}(1979)}]{ree_sem_79}
{Rees}, D.~E., \& {Semel}, M.~D. 1979, \aap, 74, 1

\bibitem[{{R{\'e}gnier} {et~al.}(2002){R{\'e}gnier}, {Amari}, \&
  {Kersal{\'e}}}]{reg_ama_02}
{R{\'e}gnier}, S., {Amari}, T., \& {Kersal{\'e}}, E. 2002, \aap, 392, 1119

\bibitem[{{R{\'e}gnier} \& {Priest}(2007)}]{reg_pri_07b}
{R{\'e}gnier}, S., \& {Priest}, E.~R. 2007, \apjl, 669, L53

\bibitem[{{Schou} {et~al.}(2012){Schou}, {Scherrer }, {Bush}, {Wachter},
  {Couvidat}, {Rabello-Soares}, {Bogart}, {Hoeksema}, {Liu}, {Duvall Jr.},
  {Akin}, {Allard}, {Miles}, {Rairden}, {Shine}, {Tarbell}, {Title}, {Wolfson},
  {Elmore}, {Norton}, \& {Tomczyk}}]{scho_sche_12}
{Schou}, J., {Scherrer }, P.~H., {Bush}, R.~I., {et~al.} 2012, \solphys, 275,
  229

\bibitem[{{Schrijver} {et~al.}(2006){Schrijver}, {Derosa}, {Metcalf}, {Liu},
  {McTiernan}, {R{\'e}gnier}, {Valori}, {Wheatland}, \&
  {Wiegelmann}}]{schr_der_06}
{Schrijver}, C.~J., {Derosa}, M.~L., {Metcalf}, T.~R., {et~al.} 2006, \solphys,
  235, 161

\bibitem[{{Schrijver} {et~al.}(2008){Schrijver}, {DeRosa}, {Metcalf}, {Barnes},
  {Lites}, {Tarbell}, {McTiernan}, {Valori}, {Wiegelmann}, {Wheatland},
  {Amari}, {Aulanier}, {Demoulin}, {Fuhrmann}, {Kusano}, {Regnier}, \&
  {Thalmann}}]{schr_08}
{Schrijver}, C.~J., {DeRosa}, M.~L., {Metcalf}, T., {et~al.} 2008, \apj, 675,
  1637

\bibitem[{{Skumanich} \& {Lites}(1987)}]{sku_lit_87}
{Skumanich}, A., \& {Lites}, B.~W. 1987, \apj, 322, 473

\bibitem[{{Sun} {et~al.}(2012){Sun}, {Hoeksema}, {Liu}, {Wiegelmann},
  {Hayashi}, {Chen}, \& {Thalmann}}]{sun_hoe_12}
{Sun}, X., {Hoeksema}, J.~T., {Liu}, Y., {et~al.} 2012, \apj, 748, 77

\bibitem[{{Thalmann} \& {Wiegelmann}(2008)}]{tha_wie_08a}
{Thalmann}, J.~K., \& {Wiegelmann}, T. 2008, \aap, 484, 495

\bibitem[{{Thalmann} {et~al.}(2008){Thalmann}, {Wiegelmann}, \&
  {Raouafi}}]{tha_wie_08b}
{Thalmann}, J.~K., {Wiegelmann}, T., \& {Raouafi}, N.-E. 2008, \aap, 488, L71

\bibitem[{{Turmon} {et~al.}(2010){Turmon}, {Jones}, {Malanushenko}, \&
  {Pap}}]{tur_jon_10}
{Turmon}, M., {Jones}, H.~P., {Malanushenko}, O.~V., \& {Pap}, J.~M. 2010,
  \solphys, 262, 277

\bibitem[{{Wheatland} \& {Metcalf}(2006)}]{whe_met_06}
{Wheatland}, M.~S., \& {Metcalf}, T.~R. 2006, \apj, 636, 1151

\bibitem[{{Wheatland} {et~al.}(2000){Wheatland}, {Sturrock}, \&
  {Roumeliotis}}]{whe_00}
{Wheatland}, M.~S., {Sturrock}, P.~A., \& {Roumeliotis}, G. 2000, \apj, 540,
  1150

\bibitem[{{Wiegelmann}(2004)}]{wie_04}
{Wiegelmann}, T. 2004, \solphys, 219, 87

\bibitem[{{Wiegelmann} \& {Inhester}(2006)}]{wie_inh_06}
{Wiegelmann}, T., \& {Inhester}, B. 2006, \solphys, 236, 25

\bibitem[{{Wiegelmann} \& {Inhester}(2010)}]{wie_inh_10}
---. 2010, \aap, 516, A107

\bibitem[{{Wiegelmann} {et~al.}(2012){Wiegelmann}, {Thalmann}, {Inhester},
  {Tadesse}, {Sun}, \& {Hoeksema}}]{wie_tha_12}
{Wiegelmann}, T., {Thalmann}, J.~K., {Inhester}, B., {et~al.} 2012, \solphys,
  67

\bibitem[{{Wiegelmann} {et~al.}(2008){Wiegelmann}, {Thalmann}, {Schrijver},
  {Derosa}, \& {Metcalf}}]{wie_tha_08}
{Wiegelmann}, T., {Thalmann}, J.~K., {Schrijver}, C.~J., {Derosa}, M.~L., \&
  {Metcalf}, T.~R. 2008, \solphys, 247, 249

\bibitem[{{Wiegelmann} {et~al.}(2010){Wiegelmann}, {Yelles Chaouche},
  {Solanki}, \& {Lagg}}]{wie_yel_10}
{Wiegelmann}, T., {Yelles Chaouche}, L., {Solanki}, S.~K., \& {Lagg}, A. 2010,
  \aap, 511, A4

\end{thebibliography}

\end{document}